# Trouble with the Lorentz Law of Force: Response to Critics

Masud Mansuripur

College of Optical Sciences, The University of Arizona, Tucson, Arizona 85721

**Abstract**. In a recent paper, we questioned the validity of the Lorentz law of force in the presence of material media that contain electric and/or magnetic dipoles. A number of authors have criticized our methods and conclusions. This paper is an attempt at answering the critics and elaborating the relevant issues in some detail.

**1. Introduction**. In a recent paper [1], we argued that a direct application of the Lorentz law of force to magnetic materials is incompatible with special relativity and with momentum conservation laws. We then suggested that the Lorentz law should be replaced by the force law first proposed by Albert Einstein and Jakob Laub in 1908 [2,3]. The Einstein-Laub force-density and its accompanying torque-density equations are

$$\boldsymbol{F}(\boldsymbol{r},t) = \rho_{\text{free}}\boldsymbol{E} + \boldsymbol{J}_{\text{free}} \times \mu_o \boldsymbol{H} + (\boldsymbol{P}\cdot\nabla)\boldsymbol{E} + (\partial\boldsymbol{P}/\partial t)\times \mu_o \boldsymbol{H} + (\boldsymbol{M}\cdot\nabla)\boldsymbol{H} - (\partial\boldsymbol{M}/\partial t)\times \varepsilon_o \boldsymbol{E}, \quad (1)$$

$$\boldsymbol{T}(\boldsymbol{r},t) = \boldsymbol{r}\times\boldsymbol{F}(\boldsymbol{r},t) + \boldsymbol{P}(\boldsymbol{r},t)\times\boldsymbol{E}(\boldsymbol{r},t) + \boldsymbol{M}(\boldsymbol{r},t)\times\boldsymbol{H}(\boldsymbol{r},t). \quad (2)$$

In the above equations, $\boldsymbol{F}$ is the force-density at point $\boldsymbol{r}$ in space and instant $t$ in time, $\boldsymbol{T}$ is the torque-density, $\rho_{\text{free}}$ and $\boldsymbol{J}_{\text{free}}$ are the local densities of free charge and free current, $\boldsymbol{P}$ and $\boldsymbol{M}$ are the local densities of polarization and magnetization, $\boldsymbol{E}$ and $\boldsymbol{H}$ are the electric and magnetic fields at $(\boldsymbol{r},t)$, and $\varepsilon_o$ and $\mu_o$ are the permittivity and permeability of free space. The system of units adopted here is *SI* (*Système international d'unités*) and the displacement and magnetic induction fields appearing in Maxwell's macroscopic equations are assumed to be $\boldsymbol{D} = \varepsilon_o \boldsymbol{E} + \boldsymbol{P}$ and $\boldsymbol{B} = \mu_o \boldsymbol{H} + \boldsymbol{M}$, respectively. The proof that the Einstein-Laub equations in conjunction with Maxwell's macroscopic equations are consistent with the conservation laws of linear and angular momentum, whereas a straightforward application of the Lorentz force and torque equations violates these laws, has been discussed in great detail in our previous publications [4-13].

The incompatibility of the Lorentz law with special relativity was demonstrated in [1] via a simple thought experiment, which is reproduced here in Fig.1. The point-charge $q$ and the magnetic point-dipole $m_o\hat{\boldsymbol{x}}'$, separated by a distance $d$ along the $z'$-axis, are stationary in the $x'y'z'$ frame. In this reference frame the particles do not interact electromagnetically, meaning that neither a force nor a torque is exerted on either particle by the electromagnetic (EM) field produced by the other. In contrast, in the $xyz$ frame, within which the particles move together at constant velocity $V$ along the $z$-axis, the dipole experiences a net torque $\boldsymbol{T} = (Vqm_o/4\pi d^2)\hat{\boldsymbol{x}}$ from the EM field of the point-charge. We argued in [1] that the existence of this torque in the $xyz$ frame, in the absence of a corresponding torque in the $x'y'z'$ frame, is proof of the inconsistency of the Lorentz law with special relativity. In contrast, the Einstein-Laub force and torque density equations yield zero force and zero torque in both reference frames, in compliance with the requirements of special relativity.

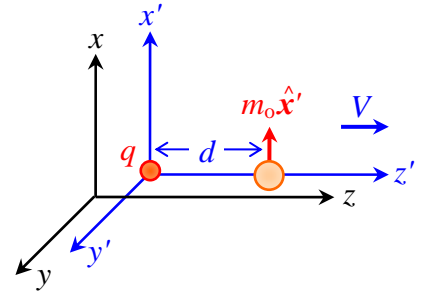

**Fig.1**. In the inertial $x'y'z'$ frame, the point-charge $q$ and the point-dipole $m_o\hat{\boldsymbol{x}}'$ are stationary. The $x'y'z'$ system moves with constant velocity $V$ along the $z$-axis, as seen by a stationary observer in the $xyz$ frame. The origins of the two coordinate systems coincide at $t = t' = 0$.

The main criticism against the above argument has been on the grounds that we had ignored the role of the so-called "hidden momentum" in our analysis of the torque experienced by the magnetic dipole, with some authors finding fault with certain other aspects of our analysis as well [14-22]. We had tried to stave off the "hidden momentum" criticism in the last paragraph of Sec.2 of [1]; in hindsight, however, it is clear that more explanation was needed. This paper is, in part, an attempt at clarifying the role of hidden momentum in the system depicted in Fig.1. We also try to address in the following sections the complaints, published as well as unpublished, of several critics, who had various disagreements with our methods and/or conclusions.

**2. Poynting's vector and Abraham's momentum density**. Let us begin by emphasizing two points that are essential for the consistency of the argument presented in the following sections.

i) In the classical theory of electrodynamics, if we treat the bound current density $J_{bound} = \mu_o^{-1}\nabla\times M$ associated with magnetization on an equal footing with the free current density $J_{free}$, we will find that the Poynting vector $S(r,t)$ is given by $E\times B/\mu_o$ and not, as is generally accepted today, by $E\times H$. The disadvantage of associating the rate of flow of EM energy with $E\times B/\mu_o$ becomes immediately clear when one considers the passage of a light pulse through a transparent slab specified by its permittivity $\varepsilon$ and permeability $\mu$. At normal incidence, one finds that a certain fraction $R$ of the optical energy is reflected at the entrance facet, while a fraction $T$ is transmitted into the slab. The sum of $R$ and $T$, however, will differ from unity because, at the entrance facet, it is the $E$ and $H$ fields that are continuous, while the $B$-field exhibits a discontinuity. One will then have to introduce the concept of "hidden energy" in order to account for the EM energy that is picked up by the induced magnetic currents at the entrance facet of the slab. In contrast, the adoption of $E\times H$ as the Poynting vector eliminates the need for hidden energy, and this has been a major argument in favor of the latter expression as the universal rate of flow of EM energy, both in free space and throughout material media. We shall assume in the following discussion that $S(r,t) = E\times H$. There is a hint here that perhaps one should draw a distinction between free and bound currents, or at least between $J_{free} + \partial P/\partial t$ on the one hand and $\mu_o^{-1}\nabla\times M$ on the other.

ii) In recent years, the Abraham momentum density $p_{EM}(r,t) = S(r,t)/c^2 = E\times H/c^2$ has gained widespread acceptance as the universal expression of EM momentum density, not only in vacuum but also in material media under all circumstances. A significant argument in support of the Abraham momentum has been provided by the Balazs thought experiment [23], where a light pulse of finite duration and finite cross-sectional area enters from vacuum into a transparent magnetic dielectric slab, traverses the length of the lab, then re-emerges into the vacuum at the exit facet. (The slab is anti-reflection coated on both ends, so the pulse enters and exits without loss of energy.) Continuity of the $E$- and $H$-fields at the various interfaces thus ensures the persistence of $S(r, t)$ as the pulse propagates, while the reduction in the group velocity by a factor $n_g$ (the group refractive index of the magnetic dielectric medium), results in a reduction of the instantaneous volume occupied by the pulse when inside the slab. As it turns out, the reduction of the pulse volume by a factor $n_g$ is precisely what is needed, in conjunction with the Abraham momentum density, to keep the center-of-energy of the system (light pulse + slab) moving at a constant speed. Any other expression for the EM momentum density will cause the slab to move at an unacceptable speed. Proponents of the alternative expression $\varepsilon_o E\times B$ for the EM momentum density, for instance, have had to *manually* add to this the "hidden momentum" density, $-\varepsilon_o E\times M$, in order to bring their arguments in line with the Balazs thought experiment [24]. In what follows, we shall assume the Abraham momentum density $\boldsymbol{E} \times \boldsymbol{H}/c^2$ for the EM field throughout the discussion. [Note that we have defined $\boldsymbol{B}$ as $\mu_o \boldsymbol{H} + \boldsymbol{M}$, which explains the difference between our hidden-momentum-density, $-\varepsilon_o \boldsymbol{E} \times \boldsymbol{M}$, and the one often encountered in the literature where $\boldsymbol{B} = \mu_o(\boldsymbol{H} + \boldsymbol{M})$, namely, $-\boldsymbol{E} \times \boldsymbol{M}/c^2$.]

**3. Current loop versus magnetic dipole**. Returning now to the "hidden momentum" objection raised by the critics, we would like to emphasize the differences that exist between an Amperian current loop associated with a magnetic dipole and ordinary current loops made up of wires, spinning rings of charge, superconducting toroids, and so forth. In connection with a magnetic dipole, traditional approach compels us to accept that an Amperian loop has a circulating current that "somehow" remains stable and charge-neutral (in the dipole's rest frame) while experiencing force and torque in the presence of a magnetic field in accordance with the Lorentz law, and exchanging energy with the local $E$-field at a rate given by $\boldsymbol{E} \cdot \boldsymbol{J}_{\text{bound}}$. To be sure, we can specify the effect of the local EM fields on dipole moments through constitutive relations [e.g., $\boldsymbol{M}(\boldsymbol{r}, \omega) = \mu_o \chi_m(\omega) \boldsymbol{H}(\boldsymbol{r}, \omega)$, where $\chi_m(\cdot)$, a function of the frequency $\omega$, is the magnetic susceptibility of a linear material], yet any internal adjustments of the Amperian loop in response to external fields, being unknown or unknowable, are generally ignored.

In contrast to the above situation with a magnetic dipole, an ordinary current loop is a known physical entity whose response to external fields can in principle be determined, either theoretically or by experimentation. Several models for such current loops have been proposed in the literature [25-30]. A nested pair of spherical shells, uniformly charged with equal and opposite surface charge densities and set in rotational motion in opposite directions around a common axis is one such model; see Fig. 2(a). Another example is a superconducting ring carrying a constant circulating current. Different models, of course, behave differently in response to an external $E$-field, but in each case it should be possible to determine the internal fields and forces that drive the circulating current. For the nested shell model of a current loop mimicking a magnetic dipole, Fig. 2(b) shows the profile of the $B$-field both inside and outside the sphere. In the absence of any magnetic material in this system, the $H$-field, aside from the trivial coefficient $\mu_o$, has the same distribution as the $B$-field.

Figure 2(c) shows a uniformly-magnetized sphere of radius $R$ placed within a uniform $E$-field produced by a pair of oppositely-charged parallel plates. In the rest frame $x'y'z'$ of the system, the $E$-field exerts neither a force nor a torque on the magnetic sphere. It is not difficult to calculate the $H$-field produced by the dipole $\boldsymbol{m}_o = (4\pi R^3/3)\boldsymbol{M}$ both inside and outside the sphere; the field lines are shown in Fig. 2(d). The integral of this $H$-field along any straight line parallel to the $x'$-axis can be readily shown to be zero. Thus the total EM (i.e., Abraham) momentum confined between the capacitor plates is exactly zero. In particular, the total EM momentum has no component along the $y'$-axis, which indicates that the field has no EM angular momentum with respect to the origin of coordinates. Seen from another inertial frame, $xyz$, in which $x'y'z'$ travels to the right along $z$ at a constant velocity $V$, the EM angular momentum of the moving system (with respect to the origin of $xyz$ in which the observer is stationary) is found to be zero once again. The bottom line is that the stationary observer in $xyz$ does *not* see any change in the EM angular momentum of the system as the system travels along $z$ at constant velocity. Consequently, there can be no torque exerted on the magnetized sphere, a conclusion that is at odds with the prediction of the Lorentz law, but consistent with that of the Einstein-Laub equations. Clearly there is no need, nor any justification, for introducing the notion of hidden momentum when dealing with magnetized bodies.

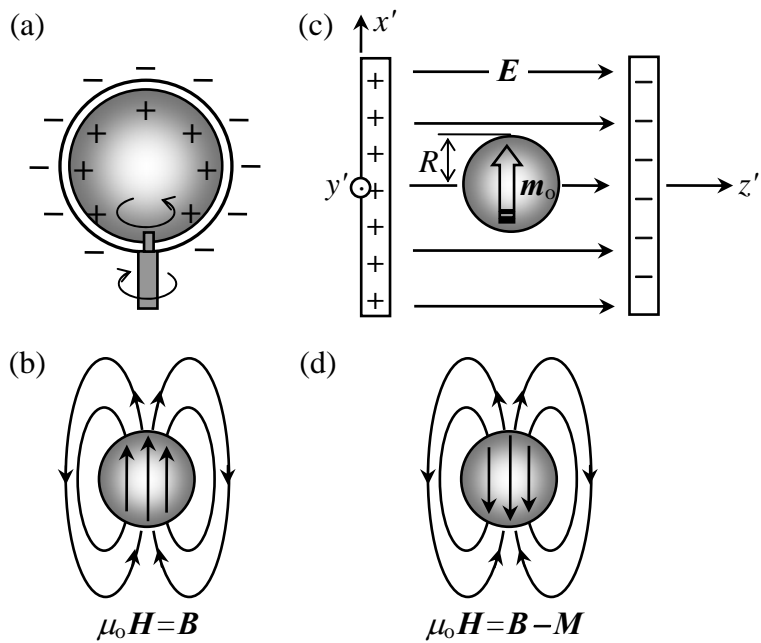

**Fig. 2**. (a) Nested pair of spherical shells, uniformly charged with equal and opposite surface charge densities and spun in opposite directions around a vertical axis, mimics a uniformly magnetized sphere. (b) Aside from the proportionality constant $\mu_o$, the $B$ and $H$ fields produced by the pair of spinning shells are the same everywhere. The uniform $H$-field inside the sphere is $H = 2M/(3\mu_o)$. (c) A uniformly-magnetized sphere of radius $R$ and magnetization $M$, immersed in a constant, uniform electric field $E$ between the plates of a capacitor. The spherical particle's dipole moment is $m_o = (4\pi R^3/3)M$. In the rest frame $x'y'z'$ of the system, both the capacitor and the magnetic dipole are stationary. (d) The $B$-field of the magnetized sphere is the same as that of the spinning shells depicted in (b). Inside the magnetized sphere, however, $H = -M/(3\mu_o)$.

Next, suppose that the pair of spinning shells of Fig. 2(a) replaces the magnetized sphere in Fig. 2(c). The magnetic fields will be the same as before, except in the region inside the sphere, where the $H$-field is now modified by the addition of $M/\mu_o$; see Fig. 2(b). The system thus has a total EM momentum $\boldsymbol{p}_{EM} = \varepsilon_o \boldsymbol{E} \times \boldsymbol{m}_o$, aligned with the $y$-axis. Since a system in which the center-of-energy is stationary cannot have a net momentum [29], the spinning shells must have an internal mechanical momentum $\boldsymbol{p}_{mech} = \varepsilon_o \boldsymbol{m}_o \times \boldsymbol{E}$ that cancels out the EM momentum. This is related to the well-known Lewis-Tolman or "right-angle lever" paradox (sometimes referred to as the Trouton-Noble paradox), whose solution has been known for at least a century [31,32]: The mechanical stresses induced within the rigid spheres in the presence of an external $E$-field do in fact give rise to the aforementioned internal mechanical momentum; see Appendices A and B. Had we chosen some other current loop to mimic a magnetic dipole, there would have been a different mechanism to produce the internal momentum (mechanical or otherwise), as has been amply elaborated upon in the literature [25-30].

With the magnetized sphere of Fig. 2(c) replaced by the pair of spinning shells, the stationary observer in the $xyz$ frame will see an EM angular momentum $\boldsymbol{L}_{EM} = z(t)\hat{\boldsymbol{z}} \times \boldsymbol{p}_{EM} = -\varepsilon_o E m_o V t \hat{\boldsymbol{x}}$, which is a linearly decreasing function of time. Conservation of angular momentum thus requires a constant torque $\boldsymbol{T} = -d\boldsymbol{L}_{EM}/dt = \varepsilon_o E m_o V \hat{\boldsymbol{x}}$ to act on the spinning shells. This torque is exactly what is needed to account for the mechanical angular momentum of the shells, $\boldsymbol{L}_{mech} = z(t)\hat{\boldsymbol{z}} \times \boldsymbol{p}_{mech} = \varepsilon_o E m_o V t \hat{\boldsymbol{x}}$, which is a linearly increasing function of time. Since we have neither a magnetic nor an electric dipole in this case – only a current loop mimicking a magnetic dipole – the Einstein-Laub force and torque become the same as the Lorentz force and torque acting on the current loop. In the $xyz$ frame, the torque $\boldsymbol{T}$ now emerges as a result of the $E$-field acting on the equal and opposite electrical charges that appear, via the Lorentz transformation of the charge-current four-vector, on the opposite faces of the moving sphere – faces that are separated by the $xz$ plane. Once again all forces, torques, momenta and angular momenta are accounted for and there is no controversy.

**4. Three-vector versus four-vector force and an alternate momentum density.** The critics assert that our treatment in [1] of the Lorentz force density as a three-vector is inappropriate, and that the correct treatment requires a four-vector analysis. This criticism would have been justified had the magnetic dipole been modeled as an *ordinary* current loop, in which case the "hidden angular momentum" would have accounted for the observed torque in the $xyz$ frame. In fact, the inclusion of hidden momentum in the analysis of the spinning spheres in the preceding section accounts for the differences between three-vector and four-vector treatments of the Lorentz force.

The critics then argue that the relativistic formalism, when properly applied, would demonstrate the compatibility of the Lorentz law of force with the fundamental tenets of special relativity. The formalism demands that we define the Poynting vector as $\boldsymbol{E}\times\boldsymbol{B}/\mu_o$ and the EM momentum density as $\varepsilon_o\boldsymbol{E}\times\boldsymbol{B}$. Under such circumstances, there appear a *hidden* energy flux, $\boldsymbol{M}\times\boldsymbol{E}/\mu_o$, and also a *hidden* momentum density, $\varepsilon_o\boldsymbol{M}\times\boldsymbol{E}$, which apparently have no physically distinct or observable impact on the energy and momentum of the system, and could as well be combined with the Poynting vector and the EM momentum density to produce an "effective" Poynting vector $\boldsymbol{E}\times\boldsymbol{H}$ and an "effective" EM momentum density $\boldsymbol{E}\times\boldsymbol{H}/c^2$. Moreover, a fraction of the Lorentz force must now be allocated to the temporal evolution of the hidden momentum, $\partial(\varepsilon_o\boldsymbol{M}\times\boldsymbol{E})/\partial t$, producing an "effective" Lorentz force density, as follows:

$$\boldsymbol{F}_{\text{effective}}(\boldsymbol{r},t) = (\rho_{\text{free}} - \nabla\cdot\boldsymbol{P})\boldsymbol{E} + (\boldsymbol{J}_{\text{free}} + \partial\boldsymbol{P}/\partial t + \mu_o^{-1}\nabla\times\boldsymbol{M})\times\boldsymbol{B} - \varepsilon_o\partial(\boldsymbol{M}\times\boldsymbol{E})/\partial t. \qquad (3)$$

We fully embrace this argument, of course, noting that it leaves the Poynting vector as $\boldsymbol{S}=\boldsymbol{E}\times\boldsymbol{H}$ and the EM momentum density as $\boldsymbol{p}_{\text{EM}}=\boldsymbol{E}\times\boldsymbol{H}/c^2$; moreover, it obviates the need for hidden entities, and shows explicitly that the force density is no longer given by the Lorentz law but by the expression in Eq.(3). This "effective" Lorentz force is in fact very similar to (but not identical with) the Einstein-Laub force; for a detailed comparison of the two force densities, see Appendix B in [33]. Since apparently there exist no observable effects of hidden energy and momentum – entities that supposedly reside within magnetic materials but have never been directly observed – our preference has always been to avoid the use of such entities in conjunction with magnetic materials. In the absence of hidden energy and hidden momentum, the incompatibility of macroscopic electrodynamics with special relativity and conservation laws can be resolved only when the "effective" Lorentz force is used to compute the force and torque exerted by EM fields on material media. To put it differently, the genius of Einstein and Laub was such that they "guessed" a form of the EM force density that was compatible with special relativity and conservation laws nearly six decades before it became apparent that the Lorentz law would accomplish the same tasks *only* with the help of hidden entities.

**5. The "effective" Lorentz force differs from the Einstein-Laub force.** The entire discussion in the preceding section would have been purely academic had the "effective" Lorentz force been in fact identical with the Einstein-Laub force. However, even in the absence of magnetism (where hidden entities play no role), the force density of the *E*-field on electric dipoles is given by $-(\nabla\cdot\boldsymbol{P})\boldsymbol{E}$ according to the Lorentz law, and by $(\boldsymbol{P}\cdot\nabla)\boldsymbol{E}$ in accordance with the Einstein-Laub law. While it can be shown that the *total* EM force and torque exerted on any given object are the same irrespective of which formulation is used, the *distribution* of force and torque densities throughout a material body will very much depend on the formulation [34,35]. The differences between the Lorentz and Einstein-Laub formulations thus become observable when *deformable* objects are subjected to EM radiation. A good example is provided by the 1973 experiments of Ashkin and Dziedzic, where the surface deformations of a transparent liquid under a focused laser beam were measured [36]. The experimental results show good agreement with theoretical analyses based on the Einstein-Laub formula [37], but are in poor agreement with calculations based on the Lorentz law [38]. We mention in passing that, to date, the vast majority of theoretical analyses in the field of radiation pressure has been confined to transparent, non-magnetic dielectrics, and based on the following restricted version of the Einstein-Laub equation:

$$\boldsymbol{F}(\boldsymbol{r},t) = (\boldsymbol{P}\cdot\boldsymbol{\nabla})\boldsymbol{E} + (\partial\boldsymbol{P}/\partial t)\times\mu_\text{o}\boldsymbol{H}. \tag{4}$$

The beauty of the Einstein-Laub formulation in conjunction with the Abraham momentum is that it eliminates the need for introducing explicit physical models for the polarization and magnetization of material media. Both linear and angular momenta are thus conserved under all circumstances, without the need to keep track of any hidden momentum. For ordinary charge and current distributions, however, the Einstein-Laub equations reduce to the Lorentz law. Once again, the linear and angular momenta of a closed system will be conserved, but a detailed knowledge of the behavior of charge carriers and their conduits in the presence of external fields may be necessary in order to predict their response and also to account for their predicted behavior.

**6. Gilbert model versus the Amperian current loop model of a magnetic dipole**. In addition to the hidden momentum argument, Griffiths and Hnizdo [17] invoke the "Lorentz force law" for a *magnetic monopole* and, referring to Namias's "Gilbert dipole" consisting of a separated pair of magnetic monopoles [26], they state that the torque on a moving magnetic dipole must be zero. Considering that their astutely-named "Lorentz force" on magnetic dipoles is equivalent to the Einstein-Laub force, it is hard to see the cause of their disagreement with our analysis. In short, we are both saying that, if the Einstein-Laub force and torque equations are accepted, there will be no torque exerted on the magnetic particle in either reference frame. We could not agree more with Griffiths and Hnizdo on this count.

**7. Consistent treatment of fields and particle trajectories**. Another criticism is that a consistent treatment of electrodynamics in the presence of material media requires the solution of coupled dynamical equations for fields and particle trajectories [22]. Let us just point out that, in our analyses, it has always been assumed that such coupled dynamical equations have *already* been solved, so that the source functions $\rho_\text{free}(\boldsymbol{r},t)$, $\boldsymbol{J}_\text{free}(\boldsymbol{r},t)$, $\boldsymbol{P}(\boldsymbol{r},t)$, $\boldsymbol{M}(\boldsymbol{r},t)$, and the radiated fields $\boldsymbol{E}(\boldsymbol{r},t)$, $\boldsymbol{H}(\boldsymbol{r},t)$ appearing in Maxwell's macroscopic equations are, in fact, the solutions of such coupled dynamical equations. Momentum conservation can then be verified in the following way:

i) Find the total EM momentum $\boldsymbol{p}_\text{EM}(t)$ of the (closed) system by integrating, over the entire space, the Abraham momentum density, $\boldsymbol{E}(\boldsymbol{r},t)\times\boldsymbol{H}(\boldsymbol{r},t)/c^2$.

ii) Calculate the total force $\boldsymbol{F}(t)$ exerted on the material media by integrating, again over the entire space, the Einstein-Laub force density $\boldsymbol{F}(\boldsymbol{r},t)$ of Eq.(1).

iii) Verify that $\boldsymbol{F}(t) + \text{d}\boldsymbol{p}_\text{EM}(t)/\text{d}t = 0$ at all times $t$.

In this way, the particle momenta make their contributions to the total momentum of the system via the force density distribution $\boldsymbol{F}(\boldsymbol{r},t)$. Similar procedures are followed to confirm the conservation of energy and of angular momentum.

**8. Point-particles and their representation by the Dirac delta-function**. One critic contends that Maxwell's macroscopic equations cannot be pushed to the limit in which the objects under consideration become point-like electric and magnetic dipoles. This critic also takes issue with our use of Dirac's delta-function to represent electric and magnetic point-dipoles. In the present section, we shall try to counter both these arguments.

There appears to be a common misconception that Maxwell's *macroscopic* equations are only approximately valid, in the sense that polarization $P(r,t)$ and magnetization $M(r,t)$ as used in these equations are meaningful only insofar as they represent average values of concentrated dipole densities, with averaging done over volumes that are large compared to atomic dimensions yet small relative to the wavelength of any electromagnetic wave that might reside within or propagate through the material medium that is host to the dipoles. This notion, however, is unnecessarily restrictive. In fact, it is possible to allow $P$ and $M$ to be arbitrary functions of the space-time coordinates $(r,t)$, without violating physical or mathematical laws. This is no different than the standard treatment of Maxwell's *microscopic* equations, in which charge and current densities, $\rho(r,t)$ and $J(r,t)$, are treated as continua, despite the fact that the underlying charge-carriers are generally acknowledged to be discrete. The material media thus make their appearance in Maxwell's macroscopic equations as $\rho_{\text{free}}(r,t)$, $J_{\text{free}}(r,t)$, $P(r,t)$ and $M(r,t)$, which entities act as sources for the electromagnetic fields $E(r,t)$ and $H(r,t)$, with $D(r,t) = \varepsilon_o E + P$ and $B(r,t) = \mu_o H + M$ appearing in the equations as composite functions of the fields and their sources. Our messy real world then mimics these beautiful and self-consistent equations in an approximate way, but not the other way around!

In conjunction with the differential form of Maxwell's equations, it is necessary, of course, for polarization and magnetization to be differentiable functions of space and time, but sharp discontinuities and lumpy or uneven distributions of matter can often be handled with the aid of such mathematical tools as the step-function and Dirac's delta-function, as well as the latter function's derivatives. We have shown elsewhere the efficacy of such techniques [39]; besides, many critics of [1] have had no difficulty accepting our use of the delta-function to represent point-charges and point-dipoles; they have performed their own calculations and arrived at exactly the same results as we have.

As beautiful as Maxwell's macroscopic equations are, they remain incomplete because they do not tell us anything about energy, momentum, force and torque. That is why we stated in [1] that we need Poynting's postulate in order to associate energy with the fields, and also to understand the exchange of energy between the fields and material media. In a similar vein, we stated that we need the Abraham postulate to associate momentum and angular momentum with the fields, and need the Einstein-Laub equations to be able to tell how momentum and angular momentum are exchanged between the fields and the material media.

We concur with Richard Feynman's statement quoted by one critic that "*We can only get a satisfactory understanding of magnetic phenomena from quantum mechanics.*" It must be added, however, that Maxwell's equations and the aforementioned auxiliary postulates – irrespective of whether one accepts the Einstein-Laub formulation or continues to use the Lorentz law – deal with magnetism as it is found in nature, without attempting to explain or to justify the magnetic behavior of material media, in general, or, that of magnetic dipoles, in particular. By the same token, Maxwell's equations and the auxiliary postulates handle electrical charge and current satisfactorily without ever being able to account for the discrete nature of charge carriers, or for the quantum-mechanical structure of atoms and molecules with which these carriers are associated.

We vehemently disagree with the critic who states that "we cannot go as far as employing mathematical tools like Dirac's delta function, and still keep $M$ and $P$ as appropriate quantities." We contend that a delta-function has a finite extent, which is as small as one may want it to be, but it is definitely *not* a zero-width function. One may imagine a small spherical particle whose diameter is much smaller than the distance over which local electromagnetic fields vary significantly. One can then use a delta-function to represent this particle and to examine its behavior in the presence of local fields. It is true that certain phenomena (e.g., radiation resistance) are difficult or impossible to study when the particle is modeled as a delta-function, but, for purposes of the present discussion, the delta-function representation is totally appropriate. Emphatically, a sufficiently small spherical magnet placed in the vicinity of a point-charge as depicted in Fig.1, will, according to the Lorentz law, experience a torque $\boldsymbol{T} = (Vqm_o/4\pi d^2)\hat{\boldsymbol{x}}$ in the $xyz$ frame. Were it not for the role played by hidden momentum in the present situation, the dipole would have exhibited observable behavior in response to the torque $\boldsymbol{T}$ exerted by the point charge $q$.

**9. Force exerted on the induced electric dipole in the moving frame**. In discussing the electric point-dipole $\boldsymbol{P}(r,t) = \gamma V \varepsilon_o m_o \delta(x)\delta(y)\delta[\gamma(z-Vt)-d]\hat{\boldsymbol{y}}$, which accompanies the magnetic dipole $m_o\hat{\boldsymbol{x}}$ in the $xyz$ frame of Fig.1, a critic has stated that the total charge of the dipole is zero. This is correct, of course, because the positive and negative charges of any dipole, when added together, cancel out. This critic then computes the Lorentz force of the *E*-field exerted by the point-charge $q$ on this dipole and finds a force $\boldsymbol{F} = -(\gamma Vm_o q/4\pi d^3)\hat{\boldsymbol{y}}$. Aside from the minus sign, he is correct once again, since at the location of the dipole, the *E*-field has a non-zero gradient along the *y*-axis, which exerts a non-vanishing force on the electric dipole. At this point the critic declares that this result is incorrect since he believes that the force on the point-dipole must be zero – ostensibly because its total charge is zero! This last statement is obviously false; the dipole indeed experiences the gradient of the *E*-field no matter how small its dimensions become, so long as its dipole moment remains constant. We mention in passing that, when the total force on the moving pair of electric and magnetic dipoles in Fig.1 is calculated, the above force on the electric dipole is cancelled out by an equal and opposite force on the magnetic dipole, resulting in a net zero force on the moving particle. The cancellation occurs not only in the Lorentz formulation, but also in the Einstein-Laub scheme.

**10. Current-current interaction**. The criticism by C. S. Unnikrishnan asserts that our analysis has failed to account for the current-current interaction. The two currents mentioned are (i) the effective current associated with the magnetic dipole (the Amperian current loop model), and (ii) the current associated with the motion of the point-charge $q$ in the $xyz$ frame. This criticism is not valid, because we have directly Lorentz-transformed (from the stationary $x'y'z'$ frame to the moving $xyz$ frame) the EM field produced by the charge $q$ at the location of the dipole. In doing so, we have fully accounted for the contribution of the moving charge $q$ to the force and torque exerted on the (moving) magnetic dipole. The magnetic field $\boldsymbol{H}(\boldsymbol{r},t)$ in Eq.(12b) of [1] is a manifestation of the motion of the charge $q$ in the $xyz$ frame. Contrary to what Dr. Unnikrishnan claims, the torque of this magnetic field on the Amperian current loop does *not* cancel out the torque exerted by the electric field of the (moving) point-charge on the induced electric dipole that arises from the motion of the magnetic dipole. Consequently, a net torque appears to act on the moving dipole when the Lorentz law of force is used to compute the force and torque experienced by the dipole.

**11. Concluding remarks**. In response to critics, we have emphasized here that the methods of calculation used in [1] are fully consistent with classical electrodynamics, and that the use of Dirac's delta-function to represent point particles is both physically and mathematically justifiable. We have argued that the critics' use of the four-vector force is equivalent to our use of the three-vector force provided that the latter calculations take hidden momentum into account. However, since in the case of magnetic materials, hidden entities are not believed to be readily observable, elimination of hidden energy and hidden momentum from the equations of classical electrodynamics will result in (i) the generally accepted expression $\boldsymbol{S}=\boldsymbol{E}\times\boldsymbol{H}$ for the Poynting vector, (ii) the Abraham momentum density $\boldsymbol{p}_{EM}=\boldsymbol{E}\times\boldsymbol{H}/c^2$ for the EM field, and (iii) the "effective" Lorentz force-density of Eq.(3). This effective force-density, while not identical with the Einstein-Laub force-density of Eq.(1), is very similar to it and, in fact, results in the same total force [and also total torque via Eq.(2)] when integrated over any finite volume of material.

In his remarks on Maxwell's equations and the early attempts to find physical models to explain the behavior of fields in vacuum, Richard Feynman wrote [40]: "*Today, we understand better that what counts are the equations themselves and not the model used to get them. We may only question whether the equations are true or false. This is answered by doing experiments, and untold numbers of experiments have confirmed Maxwell's equations. If we take away the scaffolding he used to build it, we find that Maxwell's beautiful edifice stands on its own.*" We believe the same can be said about the equations that govern the behavior of material media in the presence of EM fields, where the media are best described not in terms of free and bound charges and currents, but as spatio-temporal distributions of charge, current, polarization and magnetization.

## Appendix A: The right-angle lever paradox

An early example of the type of difficulty one encounters in relativistic analyses when force, torque, momentum and angular momentum are improperly treated is the Lewis-Tolman or the right-angle lever paradox, which is also related to the Trouton-Noble paradox [31,32]. A slightly modified version of the right-angle lever paradox, which preserves the essence of the problem but renders it somewhat easier to analyze, is shown in Fig.A1. A rigid 90º bracket (the lever) is fitted with two equal and opposite point-charges $\pm q$, then subjected to a constant uniform electric field $E_o$ at 45º to the arms of the bracket, as shown. ($E_o$ may, in fact, be the field of each charge at the location of the other, but, for the sake of generality, we shall treat it as an externally applied field.)

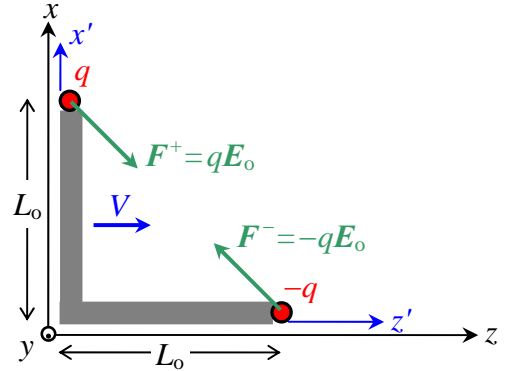

**Fig.A1**. A rigid L-shaped bracket sits at the origin of the *x′y′z′* coordinate system. Two equal and opposite point-charges, $q$ and $-q$, are affixed to the ends of the bracket and subjected to a constant uniform electric field $E_o$ at 45º to the bracket's arms. The net force and torque experienced by the bracket are zero. Seen by an observer in the *xyz* frame in which the bracket moves along *z* at the constant velocity $V$, the net force on the moving bracket remains zero, but the net torque with respect to the origin of *xyz* becomes $T=-qL_oE_o\cos 45º(V/c)^2\hat{\boldsymbol{y}}$. This torque, however, does *not* rotate the bracket, as it is needed to account for the bracket's internal angular momentum.

In the *x′y′z′* frame in which the bracket is at rest, the net force and torque on the bracket are readily seen to vanish. Next, consider the system from the perspective of an observer in the *xyz* frame, in which *x′y′z′* moves at constant velocity $V$ along the *z*-axis. The component of the *E*-field along *z* remains unchanged, while its component along *x* is multiplied by $\gamma=1/\sqrt{1-V^2/c^2}$. There will also be a *B*-field along the negative *y*-axis with magnitude $\gamma VE_o\cos 45º/c^2$. This *B*-field exerts a force $q\boldsymbol{V}\times\boldsymbol{B}=\pm\gamma qE_o(V^2/c^2)\cos 45º\hat{\boldsymbol{x}}$ on the moving charges. Thus, while the net force on each charge along $z$ is given by $\pm qE_o\sin 45°\hat{z}$ in both frames, in the $x$-direction the combined force of the $E$- and $B$-fields on each charge in $xyz$ will be reduced by a factor of $\sqrt{1-V^2/c^2}$ relative to that in the $x'y'z'$ frame. Therefore, in the $xyz$ frame, while the net force on the bracket remains zero, there appears a net torque with respect to the origin of coordinates. This is because the force $F_z$ on the $+q$ charge and the corresponding lever arm $L_o$ remain unchanged, while the force $F_x$ on $-q$ is reduced by a factor of $\sqrt{1-V^2/c^2}$ and also the arm length along $z$ suffers a Fitzgerald-Lorentz contraction by the same factor. The net torque on the bracket (with respect to the origin of the $xyz$ coordinates) is thus given by $\boldsymbol{T} = -qL_oE_o(V/c)^2\cos 45°\hat{y}$. The appearance of a torque in the $xyz$ frame in the absence of a corresponding torque in the $x'y'z'$ frame appears to be a violation of the basic tenets of special relativity.

The resolution of the paradox lies in the fact that, in the $xyz$ frame, both point-charges $\pm q$ move under the influence of a force which has a component parallel to the direction of motion. The $+q$ charge thus picks up energy from the $E$-field at a rate of $qVE_o\sin 45°$, thereby gaining momentum at a rate of $q(V^2/c^2)E_o\sin 45°$. The angular momentum associated with this linear momentum is $\boldsymbol{\mathcal{L}}(t) = -qL_o(V/c)^2E_o\sin 45°\,t\hat{y}$, whose time-rate-of-change $d\boldsymbol{\mathcal{L}}(t)/dt$ is precisely equal to the aforementioned torque $\boldsymbol{T}$.

Note that, in the $xyz$ frame, there exists a corresponding loss of energy under the influence of the $E$-field associated with the $-q$ charge. However, the acquired linear momentum in this case, lacking a lever arm with respect to the origin, does *not* contribute to the overall angular momentum. The rigidity of the bracket implies that the linear momentum gained by the $+q$ charge moves continually along the length of the bracket and feeds the $-q$ charge, thus supplying the momentum that is continuously being lost by the latter charge. In the literature, this is often treated as some sort of hidden momentum carried by the internal stresses of the bracket.

**Appendix B: Hidden momentum of a current loop in the presence of an external electric field**

Consider a ring of radius $R$ consisting of identical charged particles of charge $q$ and mass $m$, each attached to its nearest neighbors via stiff springs, as shown in Fig. B1. The ring rotates at the constant angular velocity $\omega_o$, so that each particle may be said to have a constant linear velocity $V_o = R\omega_o$ in the azimuthal direction $\hat{\phi}$. For a charge-neutral model of a magnetic dipole, we may assume the presence of a second ring, not shown in Fig. B1, with equal and opposite charges, which spins in the opposite direction and is placed immediately inside or outside the ring depicted in Fig. B1. The electrical repulsion of like charges is thus cancelled out by the attraction of opposite charges, resulting in zero electric force on each charge. (Strictly speaking, for the last statement to hold true, the charge distribution around the rings must be uniform and the two rings must occupy the exact same physical space.) When the ring is at rest, i.e., not spinning, the springs are relaxed and no force is exerted on the particles. However, once the ring is set in rotational motion, the springs stretch a little, thus exerting a net force on each particle, which accounts for the centripetal force. Each particle will now have a linear momentum $\boldsymbol{p}_o = mV_o\hat{\phi}$, with $m$ being the mass of the particle plus one half the mass of each spring which the particle shares with its nearest neighbors.

Let the total charge of the ring be denoted by $Q$. Then the charge content of a small arc of angular width $d\phi$ will be $(Q/2\pi)d\phi$. If this charge goes through a cross section of the ring in time $dt$, the current circulating around the ring will be $I = Q\omega_o/2\pi$, where $\omega_o = d\phi/dt$. Considering that the area of the loop is $A = \pi R^2$, the magnetic dipole moment of the ring will be

$$\boldsymbol{m} = \mu_o IA\hat{z} = \tfrac{1}{2}\mu_o QR^2\omega_o\hat{z}. \tag{B1}$$

The permeability $\mu_o$ of free space appears in the above expression for the magnetic dipole moment because we are defining the B-field as $\mu_o \mathbf{H} + \mathbf{M}$ rather than $\mu_o(\mathbf{H} + \mathbf{M})$. The angular momentum of the ring is readily found to be $\mathcal{M}R^2\omega_o\hat{z}$, where $\mathcal{M}$ is the ring's total mass. The ratio of the magnetic moment to the angular momentum is thus $\mu_o Q/(2\mathcal{M})$, independently of the radius $R$ and the angular velocity $\omega_o$ of the ring. For a pair of overlapping rings with opposite charges and opposite angular velocities, the magnetic dipole moments will have the same sign and will therefore be additive. The angular momenta, however, will have opposite signs and their magnitudes will have to be subtracted from each other to yield the total angular momentum. Different assignments of charge and mass to the two rings will thus result in different values for the magnetic moment and the angular momentum of the overall system.

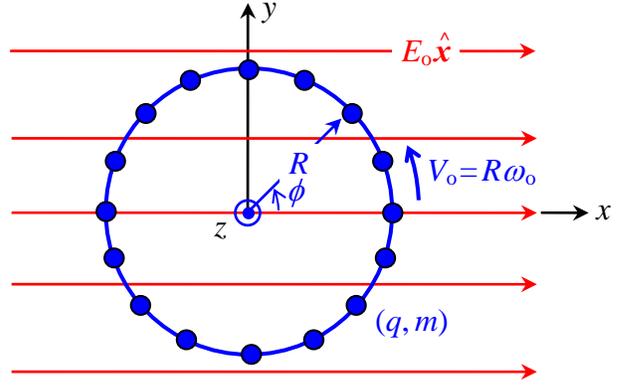

**Fig. B1**. A ring of radius $R$ consists of a large number of point-particles of mass $m$ and charge $q$, uniformly distributed around the circumference of a circle and joined together by stiff springs. The ring rotates around the $z$-axis at a constant angular velocity $\omega_o$. An identical ring (not shown) with charges $-q$ and spinning in the opposite direction is assumed to coincide with the depicted ring. This will make the pair of spinning rings charge-neutral while doubling the total magnetic moment of the system. In the presence of a constant, uniform external electric field $E_o\hat{x}$, the momentum carried by each particle at a given instant of time will be a function of the particle's azimuthal coordinate $\phi$.

In the presence of a constant uniform external field $E_o\hat{x}$, the springs adjust themselves once again, this time by an amount that depends on the position around the ring. The particle momenta now become

$$\mathbf{p}(\phi) = \mathbf{p}_o + p_1 \cos\phi\,\hat{\boldsymbol{\phi}}, \tag{B2}$$

where $\phi$ is the azimuthal angle in polar coordinates, as shown in Fig. B1. The reasoning behind the assumed form for Eq. (B2) and the relationship between $p_1$ and the system parameters will become clear in the following paragraph. Note that the change of momentum in the presence of an external $E$-field may be the result of a change in the particle's velocity or mass, or a combination of the two. In particular, the stress energy stored in the springs may account for a change of the mass $m$ associated with each particle. When the spring stiffness is high, the change in the velocity $V_o$ may be negligible, in which case the entire change in momentum must be associated with the internal stresses induced in the springs by the externally applied $E$-field.

Assuming the springs are sufficiently rigid that the presence of the $E$-field does not alter the particle velocities, the work done by the $E$-field on a given particle while traveling the distance $Rd\phi$ is given by $-qE_o\sin\phi\,Rd\phi$. This work (or energy), when divided by the square of the speed $c$ of light in vacuum, gives the change in the relativistic mass $\gamma m$ of the particle in going from $\phi$ to $\phi + d\phi$. Here, as usual, $\gamma = 1/\sqrt{1 - V^2/c^2}$. The change $dp$ in the particle momentum is, therefore, obtained by multiplying the above change in $\gamma m$ by the particle velocity $V_o = R\omega_o$, that is, $dp = -(qE_o\sin\phi\,R^2\omega_o/c^2)\,d\phi$. Recalling that $dp(\phi)/d\phi = -p_1\sin\phi$, we find

$$p_1 = qE_o R^2\omega_o/c^2. \tag{B3}$$

Next, we substitute $q = (Q/2\pi)\mathrm{d}\phi$ in the above expression for $p_1$, multiply Eq. (B2) by $\cos\phi$ to find the projection of $\boldsymbol{p}(\phi)$ on the $y$-axis, and integrate over $\phi$ from 0 to $2\pi$ to determine the total $y$-component of the ring's momentum along the $y$-axis, namely,

$$p_y = \sum_{\text{all particles}} \cos\phi\, p(\phi) = \int_0^{2\pi} [QE_\mathrm{o} R^2 \omega_\mathrm{o}/(2\pi c^2)]\cos^2\phi\, \mathrm{d}\phi = \tfrac{1}{2} QE_\mathrm{o} R^2 \omega_\mathrm{o}/c^2 = \varepsilon_\mathrm{o} m E_\mathrm{o}. \tag{B4}$$

A similar calculation for $p_x$ yields zero for the projection of the ring's momentum along the $x$-axis. The so-called "hidden momentum" of the ring in the presence of an external $E$-field is thus given by

$$\boldsymbol{p} = \varepsilon_\mathrm{o}\, \boldsymbol{m} \times \boldsymbol{E}. \tag{B5}$$

The force exerted on each particle may now be determined from Eq. (B2), as follows:

$$\begin{aligned}
\boldsymbol{F} &= \frac{\mathrm{d}\boldsymbol{p}}{\mathrm{d}t} = \frac{\mathrm{d}\boldsymbol{p}_\mathrm{o}}{\mathrm{d}t} + \frac{\mathrm{d}p_1 \cos\phi\, \hat{\boldsymbol{\phi}}}{\mathrm{d}t} \\
&= mR\omega_\mathrm{o}\frac{\mathrm{d}\hat{\boldsymbol{\phi}}}{\mathrm{d}t} + p_1\cos\phi\,\frac{\mathrm{d}\hat{\boldsymbol{\phi}}}{\mathrm{d}t} - p_1\sin\phi\,\frac{\mathrm{d}\phi}{\mathrm{d}t}\hat{\boldsymbol{\phi}} \\
&= -(mR\omega_\mathrm{o} + p_1\cos\phi)\omega_\mathrm{o}\hat{\boldsymbol{r}} - p_1\omega_\mathrm{o}\sin\phi\,\hat{\boldsymbol{\phi}}.
\end{aligned} \tag{B6}$$

The force in the $\hat{\boldsymbol{\phi}}$ direction, $F_\phi(\phi) = -p_1\omega_\mathrm{o}\sin\phi = -(R\omega_\mathrm{o}/c)^2 qE_\mathrm{o}\sin\phi$, is an attenuated version of the force $-qE_\mathrm{o}\sin\phi\,\hat{\boldsymbol{\phi}}$ exerted by the $E$-field on the charge $q$ at the azimuthal location $\phi$. The attenuation, of course, is mediated by the unequal action of the springs attached to the opposite ends of the charged particle. Also modified in the presence of the $E$-field is the centrifugal force, which, according to Eq. (B6) has an additional component $F_r(\phi) = -p_1\omega_\mathrm{o}\cos\phi = -(R\omega_\mathrm{o}/c)^2 qE_\mathrm{o}\cos\phi$. This is an inverted and attenuated version of the radial force $qE_\mathrm{o}\cos\phi\,\hat{\boldsymbol{r}}$ exerted by the $E$-field on the charge $q$ at the azimuthal location $\phi$. A corrective centrifugal force, $\boldsymbol{F}(\phi) = -[1 + (R\omega_\mathrm{o}/c)^2] qE_\mathrm{o}\cos\phi\,\hat{\boldsymbol{r}}$, must thus be supplied by whatever restraining mechanism is used to hold the ring in place (e.g., the second ring with opposite charge and opposite spin direction).

**Acknowledgements**. The author is grateful to Rodney Loudon, Tobias Mansuripur, Daniel Vanzella, Ewan Wright and Armis Zakharian for many helpful discussions.